\documentclass[aps,prl,longbibliography,twocolumn,floatfix]{revtex4-1}
\usepackage{graphicx}
\usepackage{amsmath}
\usepackage{amssymb}
\usepackage{bm}
\usepackage{times}
\usepackage[pdftex]{hyperref}
\hypersetup{colorlinks=true,linkcolor=blue,citecolor=blue,urlcolor=blue}
\usepackage{verbatim}

\renewcommand{\vec}[1]{\mathbf{#1}}
\newcommand{\var}{\operatorname{var}}
\newcommand{\calN}{{M}}

\begin{document}

\title{Absence of hyperuniformity in amorphous hard-sphere packings of
  nonvanishing complexity}
\author{M.~J.~Godfrey} \author{M.~A.~Moore}
\affiliation{School of Physics and Astronomy, University of Manchester, Manchester M13 9PL, UK}

\date{\today}

\begin{abstract}
We relate the structure factor $S(\vec{k} \to \vec{0})$ in a system of
jammed hard spheres of number density $\rho$ to its complexity per
particle $\Sigma(\rho)$ by the formula $S(\vec{k} \to \vec{0})=-1/
[\rho^2\Sigma''(\rho)+2\rho\Sigma'(\rho)]$.  We have verified this
formula for the case of jammed disks in a narrow channel, for which it
is possible to find $\Sigma(\rho)$ and $S(\vec{k})$ analytically.
Hyperuniformity, which is the vanishing of $S(\vec{k} \to \vec{0})$,
will therefore not occur if the complexity is nonzero.  An example is
given of a jammed state of hard disks in a narrow channel which is
hyperuniform when generated by dynamical rules that produce a
non-extensive complexity.
\end{abstract}
\maketitle

Systems that are hyperuniform have small long-wavelength density
fluctuations.  In other words, their structure factor $S(\vec{k})$
vanishes as the wavevector $\vec{k} \to \vec{0}$.  The structure factor
of $N$ spheres whose centers are at $\vec{r}_j$ is defined as
$S(\vec{k})= \lvert \rho_{\vec{k}}\rvert^2/N$, where the Fourier transform
of the number density $\rho_{\vec{k}}=\sum_{j=1}^N\exp(-i
\vec{k}\cdot\vec{r}_j)$.  The whole subject of hyperuniformity has been
recently  reviewed at length by Torquato \cite{torquato:18}.  In this
paper we focus on one of the most extensively studied questions
connected with hyperuniformity -- the extent to which jammed systems
of hard spheres or disks are hyperuniform.  Torquato and Stillinger
\cite{torquato:03} have made the following conjecture:\\
\textit{Any strictly jammed saturated infinite packing of identical
  spheres in $\mathbb{R}^d$ is hyperuniform.}\\
The restrictions on the packing are important.  A ``strictly'' jammed
packing is one such that each particle is firstly ``locally'' trapped
by its neighbors so that it cannot be translated while their positions
are held fixed and secondly is also ``collectively jammed'' so that no
subset of the particles can be simultaneously moved so that its
members move out of contact with one another with the remainder set
fixed.  Finally, the condition of strict jamming precludes any uniform
volume-decreasing strains of the boundary, which means that their bulk
and shear moduli are infinitely large \cite{torquato:18}.  The word
``saturated'' in the conjecture eliminates packings that contain voids
large enough to accommodate a ``rattler''~\cite{torquato:18}.  Such
voids would be a source of diffuse scattering, leading to a nonzero
limiting value for the structure factor, $S(\vec{k} \to \vec{0})\ne0$.

Many investigators have reported that the limit $S(\vec{k} \to
\vec{0})$ seems to be non-zero for jammed packings of hard spheres in
three dimensions; Refs.~\cite{ikeda:15,ikeda:17} are recent examples.
In response, Torquato has asserted that the ``difficulty of ensuring
jamming as $N$ becomes sufficiently large to access the
small-wavenumber regime in the structure factor as well as the
presence of rattlers that degrade hyperuniformity makes it virtually
impossible to test the Torquato-Stillinger jamming-hyperuniformity
conjecture via current numerical packing protocols''
\cite{torquato:18}.

Ikeda and Berthier \cite{ikeda:15}, who found that $S(\vec{k} \to
\vec{0}) \ne 0$ in their work, observed that it is ``difficult to
provide a physical explanation for the existence of the observed
deviations from strict hyperuniformity, mainly because there is no
deep physical reason to expect perfect hyperuniformity in these
systems in the first place''.  In this paper we argue for the
correctness of this view, by showing that at least in the regime where
the logarithm of the number of jammed states (a quantity usually
referred to as the complexity $\Sigma_{N,V}$) is extensive and
proportional to $N$ one can derive a simple formula for the finite
value of $S(\vec{k} \to \vec{0})$ in terms of derivatives of
$\Sigma_{N,V}$.  Our formula for $S(\vec{k}\to \vec{0})$ is the
analog of that for an equilibrium system at temperature $T$:
\begin{equation}
  S(\vec{k} \to \vec{0}) = \frac{\langle \calN^2 \rangle-\langle
    \calN \rangle^2}{\langle \calN\rangle} = \rho k_B T
  \kappa_T,
\label{equilib}
\end{equation}
where $\rho$ is the number density $N/V$, and $\kappa_T$ is the
isothermal compressibility \cite{torquato:18}; $\calN$ is the
number of particles within a large sub-volume (window) of the system
of volume $\Omega$ so $\langle \calN\rangle= \rho \Omega$.  The
name ``hyperuniformity'' means that the variance of the number of
particles in $\Omega$ is vanishingly small compared to the natural
expectation that it should be proportional to $\calN$.  Hard
sphere fluids at a temperature $T$ are not hyperuniform as their
compressibility $\kappa_T$ is finite.

We shall first re-write Eq.~(\ref{equilib}) in a more useful form for
our discussion of jammed states.  Let $p$ be the pressure in the
system.  The compressibility $\kappa _T$ is related to the second
derivative of the Helmholtz free energy $F$ of the system, $F=E- T S$,
where $E$ is the energy of the system and $S$ is its entropy:
\begin{equation}
  \kappa_T \equiv -\frac{1}{V} \left ( \frac{\partial V}{\partial p}
  \right )_{T,N}=\frac{1}{V} \frac{1}{\left({\partial^2 F/\partial
      V^2}\right)_{T,N}}.
\label{2nderiv}
\end{equation}
For a system of hard spheres in $d$ dimensions, $E=\frac12Ndk_BT$, so
that we can write $\left( {\partial^2 F/\partial V^2}\right)_{T,N}=-T
\left( {\partial^2 S/\partial V^2}\right)_{T,N}$.  Then
Eq.~(\ref{equilib}) becomes
\begin{equation}
  S(\vec{k} \to \vec{0}) = - \frac{\rho}{V}
  \frac{k_B}{\left({\partial^2 S/\partial V^2}\right)_{T,N}}.
\label{entropyrel}
\end{equation}

We turn now to generalizing Eq.~(\ref{entropyrel}) to jammed states.
We imagine that an averaging process is used.  The simplest is the
Edwards procedure of giving equal weight to all the jammed states of a
given number density $\rho$.  Let us call the complexity
$\Sigma_{N,V}$, the logarithm of the number of jammed states of $N$
spheres in a volume~$V$.  Numerical methods have been recently devised
for determining $\Sigma_{N,V}$ \cite{martiniani:16}.  When it is
extensive, it is expected to be of the form $\Sigma_{N,V} = N
\Sigma(\rho)$ as $N \to \infty$ at fixed density $\rho\equiv N/V$.  It
is then our claim that the average of the structure factor $S(\vec{k}
\to \vec{0})$ over the jammed states can be obtained from
Eq.~(\ref{entropyrel}) by making the replacement
\begin{equation}
S/N \to k_B  \Sigma(\rho).
\label{edwards}
\end{equation}
This kind of idea is clearly related to those that have been
successfully used in the theory of granular systems
\cite{edwards:89}.  In terms of derivatives of the function
$\Sigma(\rho)$, Eqs.~(\ref{entropyrel}) and (\ref{edwards}) yield
\begin{equation}
  S(\vec{k} \to \vec{0})= 
  -\frac{1}{\rho^2\Sigma''(\rho)+2\rho\Sigma'(\rho)}.
\label{final}
\end{equation}
Hyperuniformity can arise if the function $\Sigma(\rho)$ has divergent
first or second derivatives, which is only likely to happen at
exceptional values of $\rho$; we shall see later that for a system of
disks in a narrow channel this happens at the two packing fractions,
$\phi_{\textrm{min}}$ and $\phi_{\textrm{max}}$, for which the system
has crystalline order.  In general, the right hand side of
Eq.~(\ref{final}) will not be zero and there will be no
hyperuniformity.  However, if the complexity is not extensive,
Eq.~(\ref{final}) cannot be used and hyperuniformity may then occur.
Later we shall give an explicit example of such behavior for a system
of hard disks in a narrow channel for which the configuration has been
generated by a set of dynamical rules that result in a non-extensive
complexity.

The result in Eq.~(\ref{final}), obtained by analogy with the
equilibrium theory, can be derived directly for jammed systems.
Denote by $V[C_N]$ the volume of a jammed configuration $C_N$ of $N$
spheres.  Then the complexity $\Sigma_{N,V}$ is given by
\begin{equation}
  e^{\Sigma_{N,V}}= \sum_{C_N} \delta(V- V[C_N]),
\label{Sigdef}
\end{equation}
where we follow the Edwards prescription \cite{edwards:89} of giving
equal weight to each of the jammed states.  (We discuss other
averaging possibilities later.)  Often it is easier to work with
the Legendre transform of Eq.~(\ref{Sigdef}), and discuss
\begin{equation}
  Z_{N,f}= \sum_{C_N} e^{-fV[C_N]} = \int e^{\Sigma_{N,V}-fV}\,dV.
\label{Zpdef}
\end{equation} 
The reciprocal of $f$ is what Edwards and Oakeshott have referred to
as the ``compactivity'' \cite{edwards:89}.

If $\Sigma_{N,V}$ is extensive, the integral on the right-hand side of
Eq.~(\ref{Zpdef}) can be done by steepest descents.  Stationarity of
the argument of the exponential gives
\begin{equation}
  f = \frac{\partial \Sigma_{N,V}}{\partial V}
  =-\rho^2 \frac{d \Sigma(\rho)}{d \rho},
\label{pfix}
\end{equation}
where the last expression follows by writing $\Sigma_{N,V} = N
\Sigma(\rho)$ with $\rho =N/V$.  Equation~(\ref{pfix}) determines $f$
as a function of the density~$\rho$.

We note here that $\Phi(N,f)\equiv-\ln Z_{N,f}= fV-\Sigma_{N,V}$ is
the athermal analog of the Gibbs free energy.  As in thermodynamics,
one may use the extensivity of $\Phi$ to show that $\Phi=N\mu(f)$,
where $\mu=-\partial\Sigma_{N,V}/\partial N$ corresponds to the
chemical potential.

The structure factor is given by the usual expression $S(\vec{k})=
\langle\tilde\rho_{\vec{k}} \tilde\rho_{-\vec{k}}\rangle/ N$, where
$\tilde\rho_{\vec{k}}$ is the Fourier transform of
$\tilde\rho(\vec{x})=\rho(\vec{x})-\langle\rho\rangle$, the deviation
of $\rho(\vec{x})$ away from its average value.  The subtraction
removes from $S(\vec{k})$ a $\delta$-function at $\vec{k}=\vec{0}$ and
makes it possible to show that
\begin{equation}
  S(\vec{k}\to \vec{0})= \frac{\var \calN}{\langle \calN\rangle},
\label{Sfinal}
\end{equation} 
where $\calN$ is the number of particles in a window of volume
$\Omega\ll V$, placed at random in $V$, in the limit where $\Omega$
and $\langle\calN\rangle$ are large~\cite{torquato:18}.


To find the distribution of $\calN$ we adopt an approach that is
similar to the elementary one used to derive the Gibbs distribution in
statistical physics.  In a typical state drawn from the Edwards
ensemble, the number of spheres $\calN$ whose centers lie within a
randomly-placed window of volume $\Omega$ follows a certain
distribution $P(\calN)$ which we suppose to be independent of the
particular state chosen, in the limit $V\to\infty$.  We expect
$P(\calN)$ to be unchanged by further averaging over the Edwards
ensemble, and to be identical to the probability that a fixed window
of volume $\Omega$ contains $\calN$ spheres.  Provided the window is
much larger than any correlation length for the system, we can use the
extensivity of the complexity to approximate the probabilities by
\begin{equation}
  P(\calN) \propto e^{\Sigma_{N-\calN,V-\Omega}+\Sigma_{\calN,\Omega}},
\label{PSigmaSigma}
\end{equation}
where the terms in the exponent are given by expressions similar to
(\ref{Sigdef}).  Expanding $\Sigma_{N-\calN,V-\Omega}$ to first order
in $\calN$ then gives
\begin{equation}
  P(\calN) \propto e^{\Sigma_{\calN,\Omega}+\mu\calN}.
\label{GCProbs}
\end{equation}
Thus we are led to consider the partition function
\begin{equation}
  Z_{\calN,\Omega} = \sum_\calN e^{\Sigma_{\calN,\Omega}+\mu\calN}.
\label{GCensemble}
\end{equation}
The steepest-descents condition for this sum recovers the expected
$\mu=-\partial\Sigma_{\calN,\Omega}/\partial\calN$, so that by writing
$\Sigma_{\calN,\Omega}=\calN\Sigma(\rho)$, with $\rho=\calN/\Omega$,
we obtain
\begin{equation}
  \mu = -d[\rho\Sigma(\rho)]/d\rho.
\label{mueqn}
\end{equation}
In the ensemble defined by (\ref{GCensemble}), the variance of $\calN$
is given by $\var{\calN} =
\left({\partial\calN}/{\partial \mu}\right)_\Omega$.  By
using this in (\ref{Sfinal}) and setting $\calN=\rho
\Omega$ we obtain
\begin{equation}
  S(\vec{k}\to \vec{0})= 
  \frac1{\calN}
  \left(\frac{\partial\calN}{\partial \mu}\right)_\Omega =
  \frac1\rho\frac{d\rho}{d\mu}.
\label{kappaviaderiv}
\end{equation}
The result anticipated in Eq.~(\ref{final}) follows from
Eqs.~(\ref{mueqn}) and~(\ref{kappaviaderiv}).  We note that, from
Eqs.~(\ref{pfix}) and (\ref{mueqn}), the right-hand side of
(\ref{kappaviaderiv}) can also be expressed as $d\rho/df\equiv\kappa$,
a quantity which we call the ``compressibility'', by a loose analogy
with Eq.~(\ref{equilib}).

We discuss briefly the nature of the averaging procedure.  In our
treatment we have, for simplicity, considered only Edwards averages
over all the possible jammed states.  In many studies, jammed states
are produced by a quench from a fluid state at lower densities and
such quenches do not normally result in equal probabilities for all of
the states at a given packing fraction.  However, the argument leading
to Eq.~(\ref{final}) should still apply, provided the complexity
remains extensive and the complexity and structure factor are
calculated using the same ensemble of states.

\begin{figure}[htb]
  \begin{center}
    \includegraphics[width=\columnwidth]{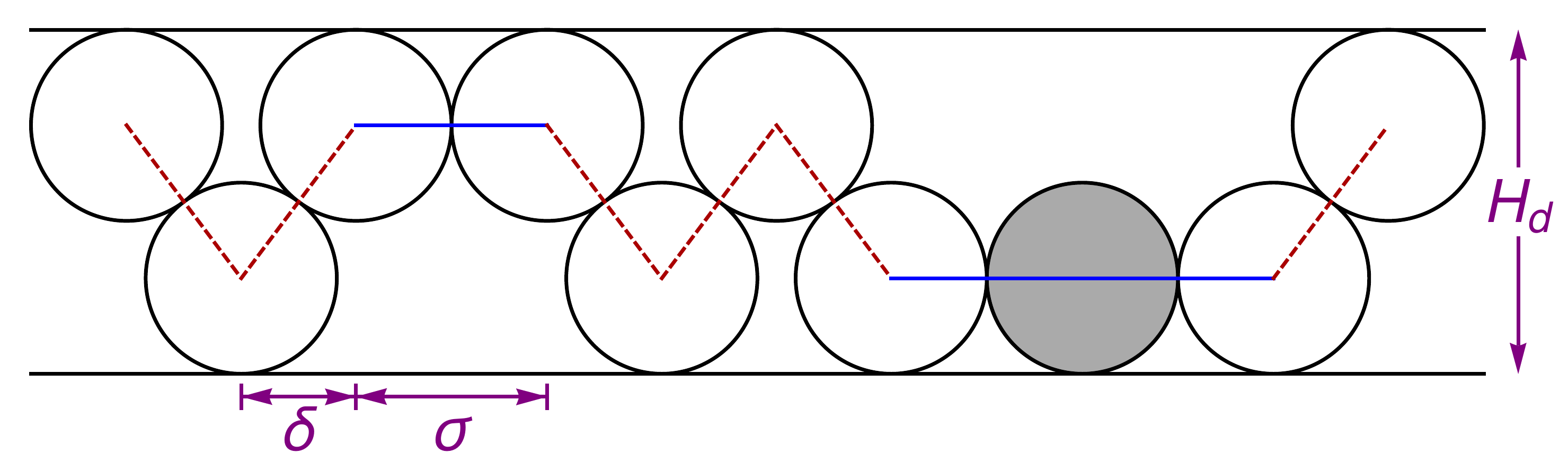}
    \caption{Local packing arrangements of hard disks of diameter
      $\sigma$ in a channel of width $H_d$.  Every disk is in contact
      with the wall and with two neighboring disks.  The shaded disk
      is not jammed.  Red dashed lines (``bonds'') join the centers of
      neighboring disks that are in contact with opposite walls and
      blue lines join the centers of disks that are in contact with
      the same wall; the latter can be regarded as ``defects'' in the
      zigzag order.  The bonds have projections $\delta$ and $\sigma$
      along the channel.}
    \label{setup}
  \end{center}
\end{figure}

We next describe a situation where one can calculate exactly both
$S(\vec{k}\to \vec{0})$ and also $\Sigma(\rho)$ for a set of jammed
states and show that they are indeed related by Eq.~(\ref{final}).
These are the jammed states in the system of disks in a narrow channel
depicted in Fig.~\ref{setup}.  The channel is of width $H_d$ and the
disks have diameter $\sigma$.  It is conventional to express densities
via the dimensionless packing fraction $\phi = \pi \rho \sigma^2/4
H_d$, where here $\rho = N/L$.  Provided $H_d/\sigma < 1+\sqrt{3/4}$,
the disks can touch only their nearest neighbors and the sides of the
channel.  The jammed states of this system are strictly jammed
saturated infinite packings.  Their only departure from the states
envisaged in the Torquato-Stillinger conjecture is that they are not
packings in the space $\mathbb{R}^d$.  Rattlers and voids large enough
to accommodate a disk only become possible for wider channels with
$H_d/\sigma \ge 1+\sqrt{3/4}$.  Locally, only a small number of
configuration types are possible: neighboring disks may touch opposite
walls or they may touch the same wall; we refer to the latter as a
``defect'' in the zigzag arrangement of the disks.  The distance
between disk centers for two disks touching the same wall is $\sigma$,
while for neighbors that touch opposite walls, the separation of their
centers, measured along the channel, is $\delta=
\sqrt{\sigma^2-(H_d-\sigma)^2}$.  Three successive disks cannot touch
the wall to form a jammed state, as the central disk is only locally
jammed and may escape by crossing the channel.  The Edwards complexity
(that is, the total number of jammed states at density $\rho$) of this
system has been determined by Ashwin and coworkers
\cite{Ashwin:11,Yamchi:15} by two methods, one based on use of the
transfer matrix and the other based on a combinatoric approach
\cite{Ashwin:11,Yamchi:15,godfrey:14}.

Let $N_D$ be the number of nearest-neighbor pairs of disks that
touch the same wall of the channel: these pairs are the defects in the
zigzag order \cite{godfrey:14}.  In terms of the concentration of
defects, $\theta \equiv N_D/N$, the density of the jammed state is
\begin{equation}
  \rho= 1/[\theta \sigma +(1-\theta) \delta].
\label{phidef}
\end{equation}
From the combinatorial approach, the complexity of the state (related
to the number of ways it is possible to place the $N_D$ defects in the
system of $N$ disks \cite{godfrey:14,krapivsky:13}) is
\begin{equation}
  \Sigma = (1- \theta) \ln(1-\theta) - \theta \ln \theta-(1-2 \theta) \ln (1-2 \theta).
\label{complexityeq} 
\end{equation}

\begin{figure}[htb]
  \begin{center}
    \includegraphics[width=\columnwidth]{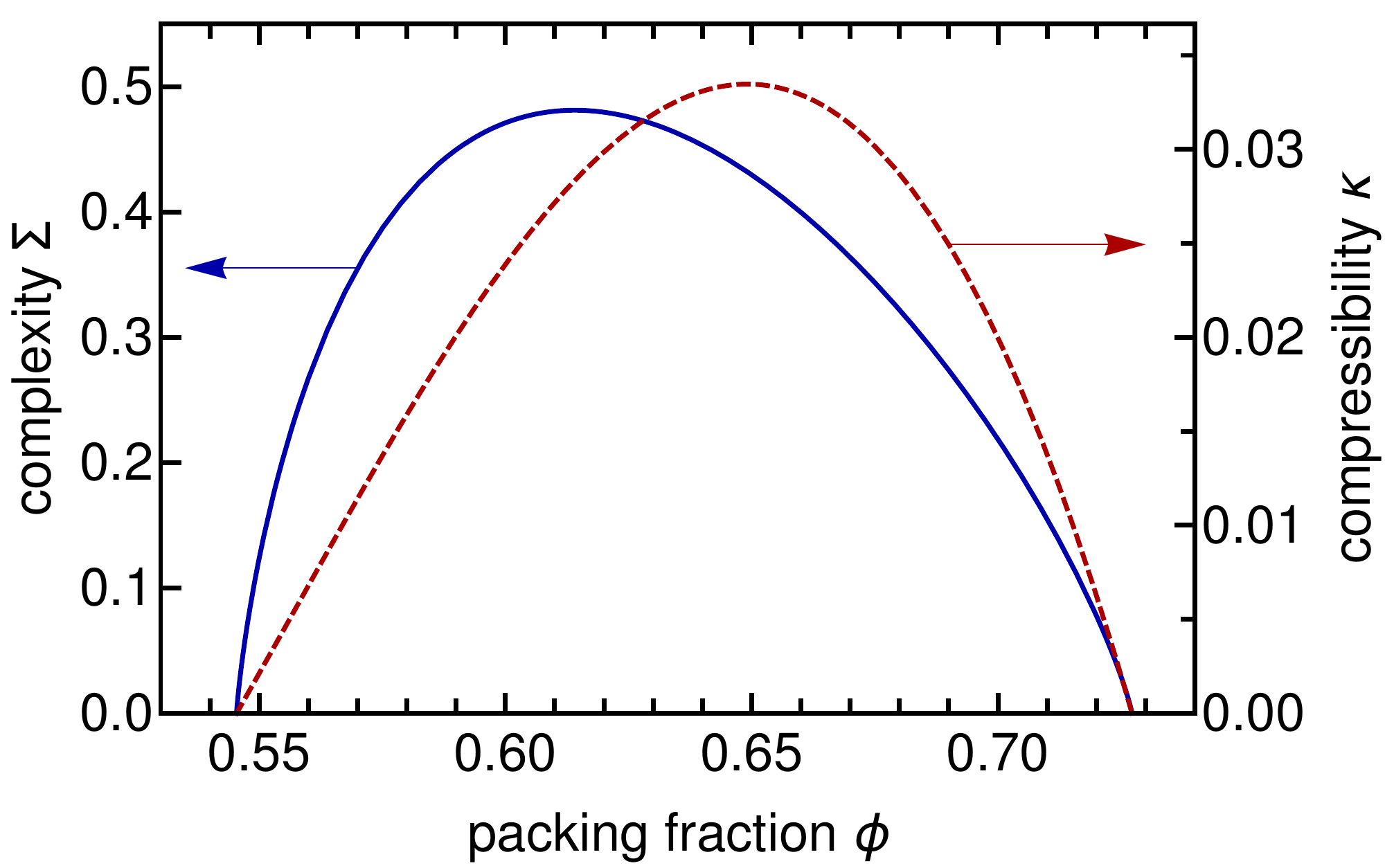}
    \caption{Complexity per particle $\Sigma$ versus packing fraction
      $\phi$ for the case $\delta/\sigma=3/5$.  Also shown is the
      ``compressibility'' $\kappa\equiv d\rho/df$; this quantity is
      equal to $S(k \to 0)$ and is nonzero for all
      $\phi_{\mathrm{min}} < \phi < \phi_{\mathrm{max}}$, implying
      that the jammed states at these densities are not hyperuniform.}
    \label{Sigma-kappa-vs-phi}
  \end{center}
\end{figure}

It is then a simple matter to use Eqs.~(\ref{phidef}) and
(\ref{complexityeq}) to calculate the complexity $\Sigma$ as a
function of the density $\rho$ or of the packing fraction $\phi$.
Figure~\ref{Sigma-kappa-vs-phi} shows the complexity as a function of
the packing fraction for the case $\delta/\sigma=3/5$.  In
Ref.~\cite{Yamchi:15} it was argued that if the jammed states are
produced by quenching from an initial fluid state, then only the
jammed states to the right of the peak in $\Sigma(\phi)$ will be
found.  Also in Fig.~\ref{Sigma-kappa-vs-phi} is plotted the
right-hand side of Eq.~(\ref{final}).  Notice that it only goes to
zero, indicating the presence of perfect hyperuniformity, at two
exceptional packing fractions, $\phi_{\mathrm{min}}$ and
$\phi_{\mathrm{max}}$; the corresponding densities are
$\rho_{\mathrm{min}}=2/(\sigma+\delta)$ and
$\rho_{\mathrm{max}}=1/\delta$.  The case of $\phi_{\mathrm{max}}$
arises when no defects are present, i.e. for $\theta=0$.  In this case
the jammed state has crystalline order and the system is hyperuniform.
On the other hand $\phi_{\mathrm{min}}$ arises when $\theta=1/2$.
This jammed state corresponds to a loosely-packed crystalline
structure.

As the starting point for the calculation of the structure factor, we
first used the transfer matrix method to obtain the partition function
in the $(N,f)$ ensemble, Eq.~(\ref{Zpdef}).  In (\ref{Zpdef}), the
volume function $V[C_N]$ becomes the total length of a configuration
of disks, $L=\sum_{i=1}^{N-1} \sigma_{i,i+1}$, where
$\sigma_{i,i+1}=\sigma$ for neighboring disks, numbered $i$ and $i+1$,
that form a defect and $\sigma_{i,i+1}=\delta$ otherwise.  An explicit
form for the resulting transfer matrix has been given in
Ref.~\cite{Yamchi:15}.  Let the largest eigenvalue of the transfer
matrix be $\lambda_1(f)$, so that $Z_{N,f}\approx\lambda_1^N$ for
large~$N$.  Then $\Sigma_{N,L}-f L= N \ln\lambda_1$.  As in
Eq.~(\ref{pfix}), $f$ can be related to $\rho$ by the steepest descent
condition $f= \partial\Sigma_{N,L}/\partial L$.  Intuitively, $f$ can
be understood as a force that is applied by a piston at the end of the
channel; $f$ can be varied to select jammed states of a particular
length $L$.

\begin{figure}[htb]
  \begin{center}
    \includegraphics[width=\columnwidth]{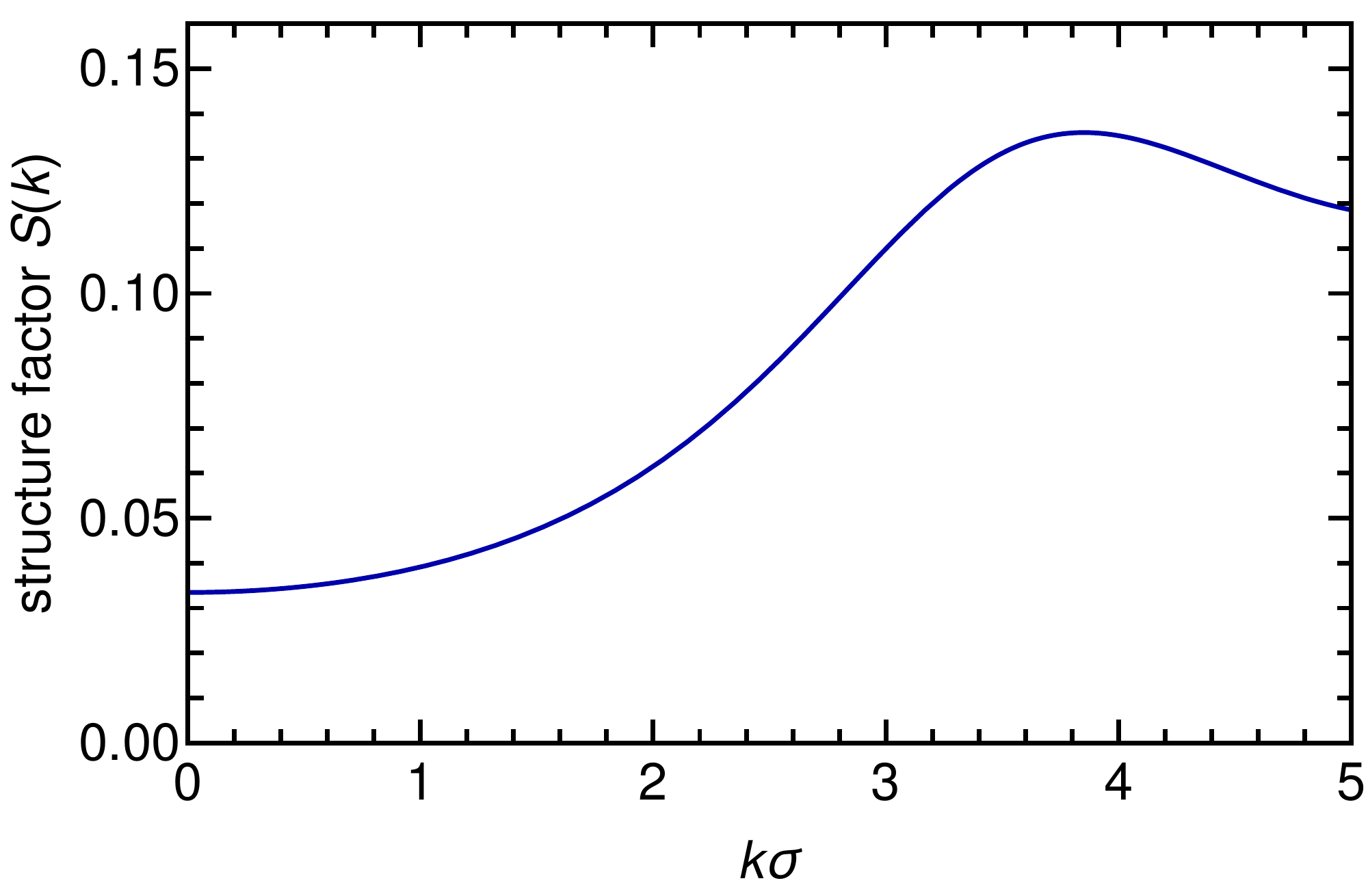}
    \caption{Structure factor $S(k)$ versus wavenumber $k$ for
      $\delta/\sigma=3/5$ at a packing fraction $\phi=0.649$, which
      corresponds to the maximum in the compressibility shown in
      Fig.~\ref{Sigma-kappa-vs-phi}.  For small $k$, $S(k) \to
      {\mathrm{const.}}+\mathcal{O}(k^2)$.}
    \label{expandedstructurekzero}
  \end{center}
\end{figure}

The structure factor $S(\vec{k})$ can be calculated in closed form
from the transfer-matrix solution by the method explained (for a
thermal problem) in Ref.~\cite{robinson:16}.  We will consider only
the case where $\vec{k}$ is in the direction of the channel, as the
main interest is the form of $S(k)$ for $k \to 0$.
Figure~\ref{expandedstructurekzero} shows $S(k)$ for small values
of~$k$.  It can be shown analytically to be of the form
$S(k)=\text{const.}+\mathcal{O}(k^2)$, where the constant is exactly
that which would have been obtained from the complexity via
Eq.~(\ref{final}).  Expressed in terms of $\rho$, the small-$k$ limit
of $S(k)$ is given by
\begin{equation}
  S(k\to0) =
  2\,\frac{(\rho\sigma-1)(1-\rho/\rho_{\mathrm{max}})(\rho/\rho_{\mathrm{min}}-1)}{\rho(\sigma-\delta)},
\end{equation}
which vanishes for $\rho=\rho_{\mathrm{min}}$ and
$\rho_{\mathrm{max}}$, corresponding to $\phi=\phi_{\mathrm{min}}$ and
$\phi_{\mathrm{max}}$ in Fig.~\ref{Sigma-kappa-vs-phi}.

The results shown in Fig.~\ref{expandedstructurekzero} were calculated
for packing fraction $\phi=0.649$ with the ratio
$r=\delta/\sigma=3/5$.  Note that if $r=p/q$ with $p$ and $q$ coprime
integers, $S(k)$ will have $\delta$-function peaks for $k=2\pi
mq/\sigma$, where $m\ne0$ is an integer.  The first of these is at
$k=10\pi/\sigma$, which lies outside the range plotted in
Fig.~\ref{expandedstructurekzero}.
 
A quite different situation arises if the total complexity is not
extensive.  An example of such a case is the Fibonacci quasicrystal
\cite{lu:86,buczek:05,oguz:17}, which can be realized as a particular
jammed state of our system of disks in a narrow channel system.  Let
$A$ denote a diagonal (red) bond between neighboring disks and $B$ a
horizontal (blue) bond in Fig.~\ref{setup}.  Thus the configuration
shown in Fig.~\ref{setup} is $AABAAABBA$.  The Fibonacci quasicrystal
is grown according to the following deterministic dynamical rule
\cite{lu:86}
\begin{equation}
A \to AB; \hspace{1cm} B \to A.
\label{dynrule}
\end{equation}
This generates the sequences
\begin{equation}
 AB \to ABA \to ABAAB \to ABAABABA \to \cdots,
\nonumber
\end{equation}
in which the numbers of $A$ and $B$ bonds in each block are
consecutive members of the Fibonacci sequence $1$, $1$, $2$, $3$, $5$,
$8$, \dots.  The dynamical rules do not allow the generation of the
configuration in Fig.~\ref{setup}, which contains a $BB$ sequence and
is not jammed.  For the limit when the final block is infinitely long,
the numbers of $A$ and $B$ are such that $N_{A}/N_B \equiv (1-\theta)/
\theta=(1+\sqrt{5})/2$, so that $\theta=(3-\sqrt5)/2$.  This value of
$\theta$ corresponds to a point to the left of the maximum of the
complexity versus $\phi$ plot in Fig.~\ref{Sigma-kappa-vs-phi}, for
which the maximum corresponds to $\theta= 1/2 -\sqrt5/10$
\cite{Ashwin:11}.  The structure factor of the infinite sequence has
been studied in Ref.~\cite{buczek:05} and is that of a quasicrystal.
One can construct windows of length $R$ within the sequence and
average over these windows: the number of distinct windows increases
as $\sim \exp(\Sigma_R)$.  (This connection between the complexity
$\Sigma_R$ and the number of distinct windows has been studied in some
detail by Kurchan and Levine \cite{kurchan:11}.)  The complexity
$\Sigma_R$ is not extensive: for a window of length $R$ in the
infinite sequence, the number of distinct windows does not exceed $R$
\cite{lu:86}, so $\Sigma_R < \ln R$.  It can also be shown that the
system is hyperuniform \cite{oguz:17}: the variance in the number of
disks $\calN$ in a window of length $R$ grows no faster than $\ln R$,
so the ratio $\var{\calN}/\langle\calN\rangle\to 0$ for $R\to\infty$,
implying that the Fibonacci quasicrystal is hyperuniform.

In conclusion, we have shown that hyperuniformity will not arise in a
jammed state of hard spheres if the complexity is extensive.  The
$\vec{k}\to\vec{0}$ limit of $S(\vec{k})$ is related to the complexity
by Eq.~(\ref{final}).  We have illustrated this for a simple system
for which both the complexity and structure factor could be determined
exactly.  If the jammed state is generated by dynamical rules that
generate a non-extensive complexity, as in our Fibonacci example, then
hyperuniformity can arise.

\begin{acknowledgments}
We should like to thank Ludovic Berthier and Salvatore Torquato for
helpful correspondence.
\end{acknowledgments}

\bibliography{refs}

\end{document}